\documentclass[amsmath,amssymb,superscriptaddress,nobalancelastpage,preprint]{revtex4-1}

\usepackage{graphicx}
\usepackage{varioref}
\usepackage{xr-hyper}
\usepackage{xcolor}
\usepackage{hyperref}
\hypersetup{colorlinks,linkcolor=blue,urlcolor=blue,citecolor=blue}
\usepackage{ulem}
\usepackage{braket}

\usepackage{xcolor}

\begin{document}

\title{Surface structure of the 3$\times$3-Si phase on Al(111), studied by the multiple usages of positron diffraction and core-level photoemission spectroscopy}

\newcommand{\AffISSP}{\affiliation{Institute for Solid State Physics (ISSP), The University of Tokyo, Kashiwa, Chiba 277-8581, Japan}}
\newcommand{\AffJAEA}{\affiliation{Advanced Science Research Center, Japan Atomic Energy Agency, 2-4 Shirakata, Tokai, Naka, Ibaraki 319-1195, Japan}}
\newcommand{\AffTottori}{\affiliation{Department of Engineering, Tottori University, Tottori,  Tottori, 680-8550, Japan}}

\newcommand{\AffNIFS}{\affiliation{National Institute for Fusion Science, 322-6 Oroshi-cho, Toki, Gifu 509-5292, Japan}}
\newcommand{\AffTokyoINST}
{\affiliation{Laboratory for Chemistry and Life Science, Institute of Innovative Research, Tokyo Institute of Technology, 4259 Nagatsuta-cho, Midori-ku, Yokohama, Kanagawa 226-8501, Japan}}
\newcommand{\AffTamkang}{\affiliation{Department of Physics, Tamkang University, No.151, Yingzhuan Rd., Tamsui Dist., New Taipei City 251301, Taiwan}}
\newcommand{\AffTohoku}{\affiliation{Institute of Multidisciplinary Research for Advanced Materials, Tohoku University, Sendai 980-8577, Japan}}

\author{Yusuke~Sato}\AffISSP
\author{Yuki~Fukaya}\AffJAEA
\author{Akito~Nakano}\AffTottori\AffNIFS
\author{Takeo~Hoshi}\AffISSP\AffNIFS
\author{Chi-Cheng~Lee}\AffTamkang
\author{Kazuyoshi~Yoshimi}\AffISSP
\author{Taisuke~Ozaki}\AffISSP
\author{Takeru~Nakashima}\AffTokyoINST
\author{Yasunobu~Ando}\AffTokyoINST
\author{Hiroaki~Aoyama}\AffTohoku
\author{Tadashi~Abukawa}\AffTohoku
\author{Yuki~Tsujikawa}\AffISSP
\author{Masafumi~Horio}\AffISSP
\author{Masahito~Niibe}\AffISSP
\author{Fumio~Komori}\AffISSP
\author{Iwao~Matsuda}\AffISSP

\date{\today}



\begin{abstract}
The structure of an Al(111)3$\times$3-Si surface was examined by combining data from positron diffraction and core-level photoemission spectroscopy. Analysis of the diffraction rocking curves indicated that the overlayer had a flat honeycomb lattice structure. Simulations of Si core-level spectra calculated via the first-principles indicated that one of the Si atoms in the unit cell was replaced by an Al atom. The surface superstructure was thus a two-dimensional layer of Al-embedded silicene on Al(111).
\end{abstract}

\maketitle

\section{Introduction}



Two-dimensional (2D) materials have been a central issue in materials science because of the emergence of physical or chemical phenomena that are specific to its dimensionality, with promising applications in nanotechnology\cite{MatsudaTextbookp1}. A Xene is a 2D material named after the elemental atomic sheet\cite{Xenes2022}, such as graphene (C)\cite{graphene2022,graphene2023,graphene2019}, silicene (Si)\cite{silicene2018,silicene2016}, or borophene (B)\cite{MatsudaTextbookp2}. Diversity in Xene research was triggered by the discovery of silicene on the Ag(111) surface\cite{Vogt2012,Takagi}. The surface structure is fundamental information for a material, and that of the silicene layer was determined by positron diffraction. It formed a honeycomb lattice with out-of-plane buckling\cite{Fukaya2013}. The electronic structure was expected to show Dirac cones at the K point in the 2D Brillouin zone, but the actual band mappings via photoemission spectroscopy revealed different positions in momentum (wavenumber) space\cite{Paolo1,Paolo2,Feng1,Feng2}. These investigations have indicated that Xene properties differ between free-standing layers and those on surfaces. Such external effects have been reported for various combinations of overlayers and substrates\cite{Feng2017,Mathis}. Detailed analyses of Xene structures and electronic states are still required for specific surfaces before a universal understanding is achieved.

On Al(111) surfaces, it has been reported that Si deposition induces the formation of a long-range ordered 2D phase, or the Al(111)3$\times$3-Si surface\cite{Jona,Munoz1,Munoz2,Sassa2020,Sato2020}. The ordered phase was found at a Si coverage of 0.4 $\rm \sim$ 0.8 ML\cite{Jona,Munoz1,Munoz2,Sassa2020}, where 1 ML corresponds to the surface atomic density of Al(111), which is $\rm 1.42 \times 10^{19} atoms/cm^{2}$. The surface band structure was probed via angle-resolved photoemission spectroscopy and consistently matched the band diagram calculated for a honeycomb lattice Si layer (silicene) on Al(111)\cite{Sato2020}. However, a chemical analysis of the surface via Si 2$p$ core-level photoemission spectroscopy indicated Si environments that were much more complex than a simple silicene layer, and the "kagome-like silicene" structure model was proposed \cite{Sassa2020}. These experiments have implied that this Xene system has unique properties, but no surface structure analysis has been reported.

Here, we formed the Al(111)3$\times$3-Si surface and conducted structure analyses via multiple experiments using positron diffraction and core-level photoemission spectroscopy. Data-driven simulations and the first-principles calculations indicated that it had a flat honeycomb lattice arrangement with  Al atoms embedded in the overlayer. The Al-embedded silicene model described the experimental results when the Si coverage was 7/9 ML. The band calculations also reproduced the photoemission band diagram reported previously\cite{Sato2020}. The Al(111)3$\times$3-Si surface is a unique silicene system where the layer has partially embedded substrate atoms, showing an intriguing external effect of the surface. This work thus suggests possible doping of Xenes by external atoms, which may lead to novel functional layers.

\begin{figure}[htp]
\centering
\includegraphics[width=8 cm]{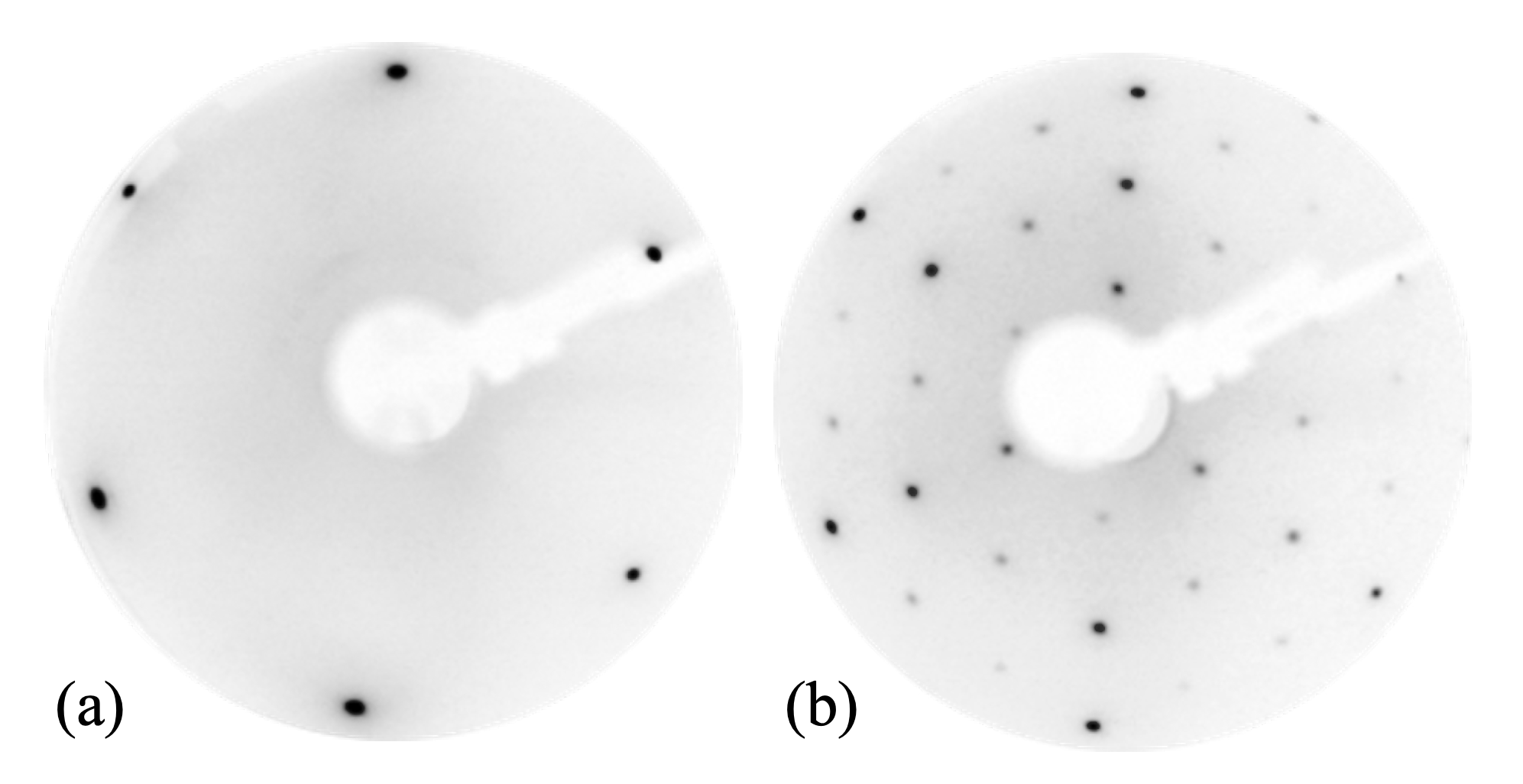}
\caption{Low-energy electron diffraction patterns acquired with 88.2 eV electrons. (a) A pristine Al(111) surface, and (b) the Al(111)3$\times$3-Si surface\cite{Sato2020}, .}
\label{f1}
\end{figure}

\section{Experiments and simulations}
\subsection{Sample preparation}
The 3$\times$3 phase was prepared via Si deposition on the Al(111) substrate in an ultrahigh vacuum chamber that had a base pressure of 2 $\times$ $\rm 10^{-10}$ mbar. A clean Al(111) surface was prepared via cycles of $\rm Ar^{+}$ sputtering and annealing at 670 K. A clean and ordered surface was confirmed by elemental analyses via X-ray photoelectron spectroscopy and electron (positron) diffraction patterns, respectively. Here, various diffraction methods were used to characterize the sample surface: low-energy electron diffraction (LEED), reflection high-energy electron diffraction, and total-reflection high-energy positron diffraction (TRHEPD). Figure 1(a) shows a LEED pattern of a pristine Al(111) surface. Si was then deposited on this Al(111) substrate at 350 K. The Si source was based on sublimation of a Si wafer via Joule heating. Formation of the Al(111)3$\times$3-Si phase was checked with a electron diffraction pattern, as shown in Fig. 1(b). The sample surface was then transferred $in$ $situ$ to perform positron diffraction and core-level spectroscopy.

\subsection{Positron diffraction}
The Al(111)3$\times$3-Si structure was examined with TRHEPD \cite{MatsudaTextbookp1,Fukaya2016,Fukaya2013,Fukaya2021,Fukaya2012p1,Fukaya2012p2,Fukaya2012p3} performed at the Slow Positron Facility at KEK\cite{Wada2012}. Details of the experimental setup were reported previously\cite{Fukaya2019}. The energy of the incident positron beam was 10 keV, and rocking curves and spot intensities as a function of glancing angle of the beam were measured by rotating the sample holder up to $\rm 6^{\circ}$ with step sizes of $\rm 0.1^{\circ}$. The surface structure was determined by searching for a model that had good agreement between experimental and simulated rocking curves. It was also quantitatively judged by a R-factor with the minimum value for the most appropriate structure model\cite{Fukaya2016,Fukaya2013,Fukaya2021,Fukaya2012p1,Fukaya2012p2,Fukaya2012p3}. The R-factor (R) was defined by:

\begin{equation}
R = \sqrt{\sum_{\theta} (I^{exp}_{\theta} - I^{sim}_{\theta})^2}.
\end{equation}

where $I^{exp}_{\theta}$ and $I^{sim}_{\theta}$ are the normalized experimental and simulated intensities at the glancing angle ($\theta$), respectively\cite{Fukaya2004}.

The data-driven analyses used TRHEPD results taken under the one-beam condition, and simulation results obtained with the 2DMAT software package\cite{Motoyama2022}. Bayesian inference was performed via the massively parallel Monte Carlo method on the "Fugaku" supercomputer. 

\subsection{Core-level photoemission spectroscopy}
Core-level photoemission spectroscopy measurements were performed at a vacuum ultraviolet photoemission beamline (Elettra, Trieste). Details were described previously\cite{Paolo1,Paolo2,MatsudaTextbookp1}. The beamline covered a photon energy range of 20 eV to 750 eV, and the base pressure was 1 $\times$ $\rm 10^{-11}$ mbar. Simulations of the spectroscopy were performed with OpenMX\cite{Ozaki2003,Ozaki2004,Ozaki2005,Lejaeghere2016,Ozaki2017,OPENMX}. In the calculations, 3,3,2 optimized radial functions were allocated for Al $s$, $p$, and $d$ orbitals with a cutoff radius of 7 Bohr. For Si atoms, 2, 2, 1 optimized radial functions were allocated for $s$, $p$, and $d$ orbitals with a cutoff radius of 7 Bohr. A Perdew-Burke-Ernzerhof generalized gradient approximation was used as the exchange correlation functional\cite{Perdew1996}. In the spectral simulations, both atomic orbitals and pseudo-potentials included contributions from the 2$p$ orbital. The energy cutoff and the energy convergence criterion were 220 Rydberg and 1.0$\times$ $\rm 10^{-8}$ Hartree, respectively. For the wavevector space, a 6$\times$6$\times$1 grid was used. In the calculations, an Al(111) slab of seven layers was used, where one side was covered with the Si(Al) surface layer. The slabs were repeatedly separated by a vacuum at a distance of 17 $\rm \AA$ and facing the bare and overlayer surfaces. The binding energy of each core-level was obtained by calculating the total energy difference before and after creation of the core-hole state\cite{Ozaki2017}.

\subsection{Electronic structure calculations}
Calculations of structural optimizations and band dispersion curves were performed using OpenMX\cite{Ozaki2003,Ozaki2004,Ozaki2005,Lejaeghere2016,Ozaki2017,OPENMX} with the same conditions used for the spectral simulations described above for optimized radial functions, a cutoff radius, and an exchange correlation functional. The energy cutoff was 300 Rydberg, while the energy convergence criterion was 1.0$\times$ $\rm 10^{-8}$ Hartree. A grid of 6$\times$6$\times$1 was used for the wavevector space. In the calculations, an Al(111) slab of five layers was adopted, where one side was covered with the Si(Al) surface layer. The slabs were repeatedly separated by vacuum at a distance of 23 $\rm \AA$. Structural optimization was performed by minimizing the force upon each atom. At the bottom of the Al(111) slab, the coordinates of the atoms were fixed, while those of the other atoms were relaxed. The force convergence criterion was 1.0 $\times$ $\rm 10^{-4}$ Hartree/Bohr. After the convergence of energy and force was achieved, the total energy of the system was calculated to assess the system stability. To simulate the 2D slab, an effective screening medium method was applied in which a semi-infinite vacuum was inserted at the bottom and top of the system\cite{Otani2006,Sugino2007,Otani2008,Ohwaki2012}.

\begin{figure}[htp]
\centering
\includegraphics[width=6 cm]{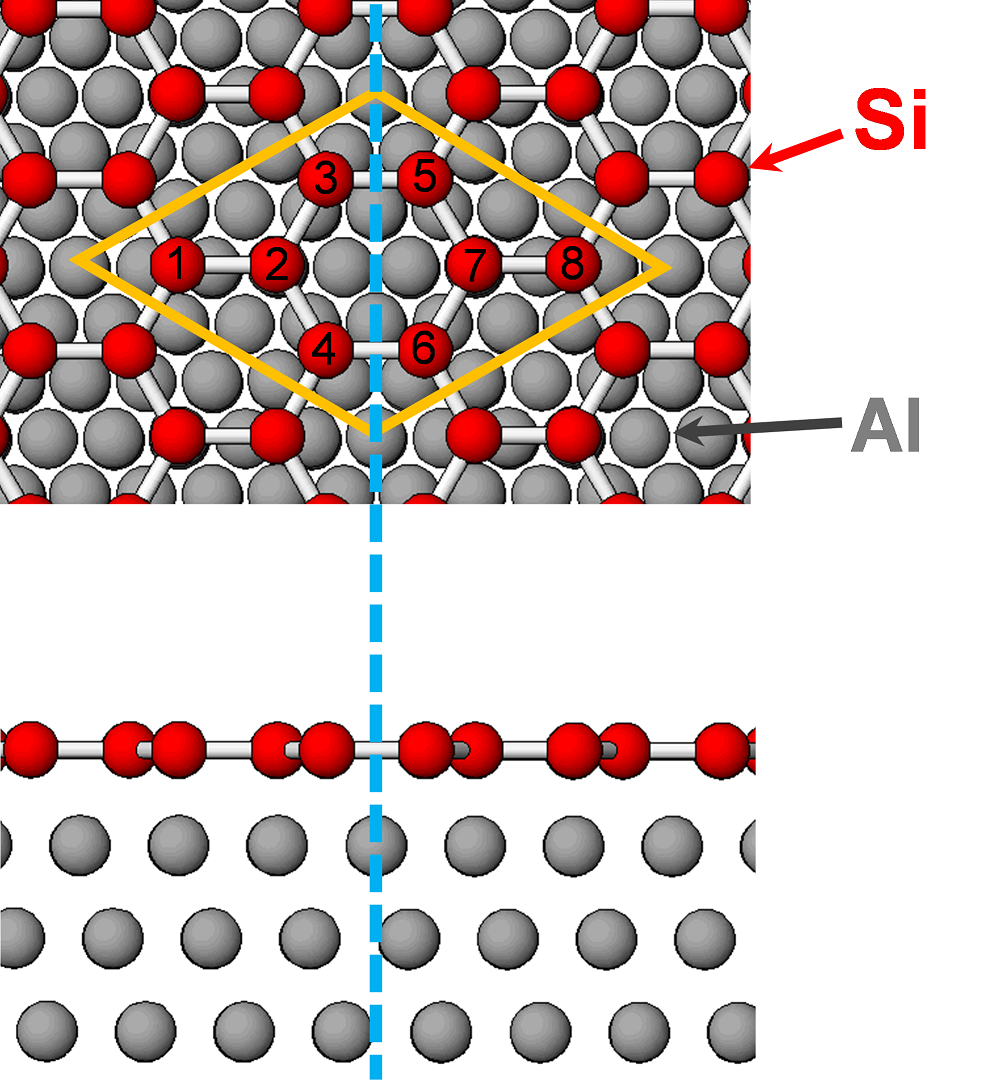}
\caption{Schematic of the flat silicene model. Si atoms are labeled with numbers 1 to 8. A 3$\times$3 unit cell is depicted.}
\label{f2}
\end{figure}

\section{Results and Discussion}
\subsection{Positron diffraction}
TRHEPD was conducted to investigate the structure of the Al(111)3$\times$3-Si surface. TRHEPD rocking curves were measured at $\rm 13^{\circ}$ off from the [$11\bar{2}$] direction (one-beam condition), and along the [$11\bar{2}$] and [$1\bar{1}$0] directions. Under the one-beam condition, the specular (00) spot intensities depended mainly on the surface-normal components (z) of atomic positions because simultaneous reflections in the surface-parallel plane were sufficiently suppressed\cite{Ichimiya1987}. In the TRHEPD analyses, a surface structure was determined by searching for a model that had agreements between simulated and experimental curves\cite{Fukaya2016,Fukaya2013,Fukaya2021,Fukaya2012p1,Fukaya2012p2,Fukaya2012p3}. A honeycomb lattice of Si atoms, shown in Fig. 2, was adopted as an initial model because it had previously described the Si layer (silicene) on Ag(111) and the Ge layer (germanene) on Al(111)\cite{Fukaya2016,Fukaya2013}.

\begin{figure}[htp]
\centering
\includegraphics[width=6.5 cm]{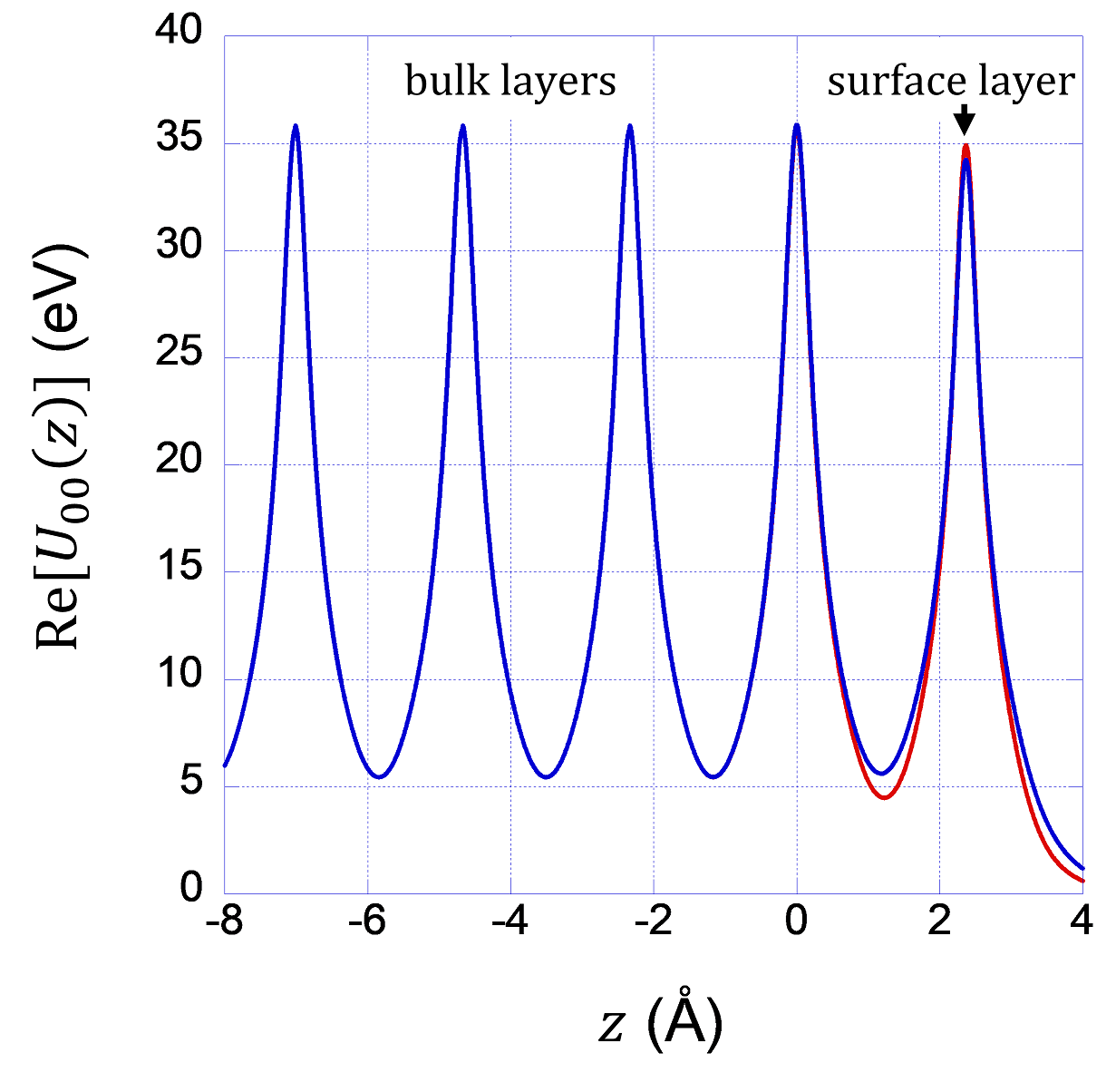}
\caption{Real parts of scattering potentials, Re[$\rm U_{00}$], averaged in the surface-parallel plane for the Si honeycomb lattice layer on Al(111) (red) without or (blue) with Al substitution. Bulk and surface layers are indicated by labels.}
\label{f3}
\end{figure}

When performing the surface analysis via positron diffraction, it is important to note that Al and Si are located next to each other in the periodic table and thus have similar scattering potentials for a positron beam. This was demonstrated by the calculations shown in Fig. 3. The scattering event can be evaluated by the positron scattering potential. The scattering potential at the Si surface layer appeared nearly identical without or with Al substitution. This indicated that the surface structure model created uncertainty regarding embedded Al atoms. However, the structure parameters of the Si(Al) surface layer can be elaborated by considering only Si atoms.  

\begin{figure}[htp]
\centering
\includegraphics[width=7 cm]{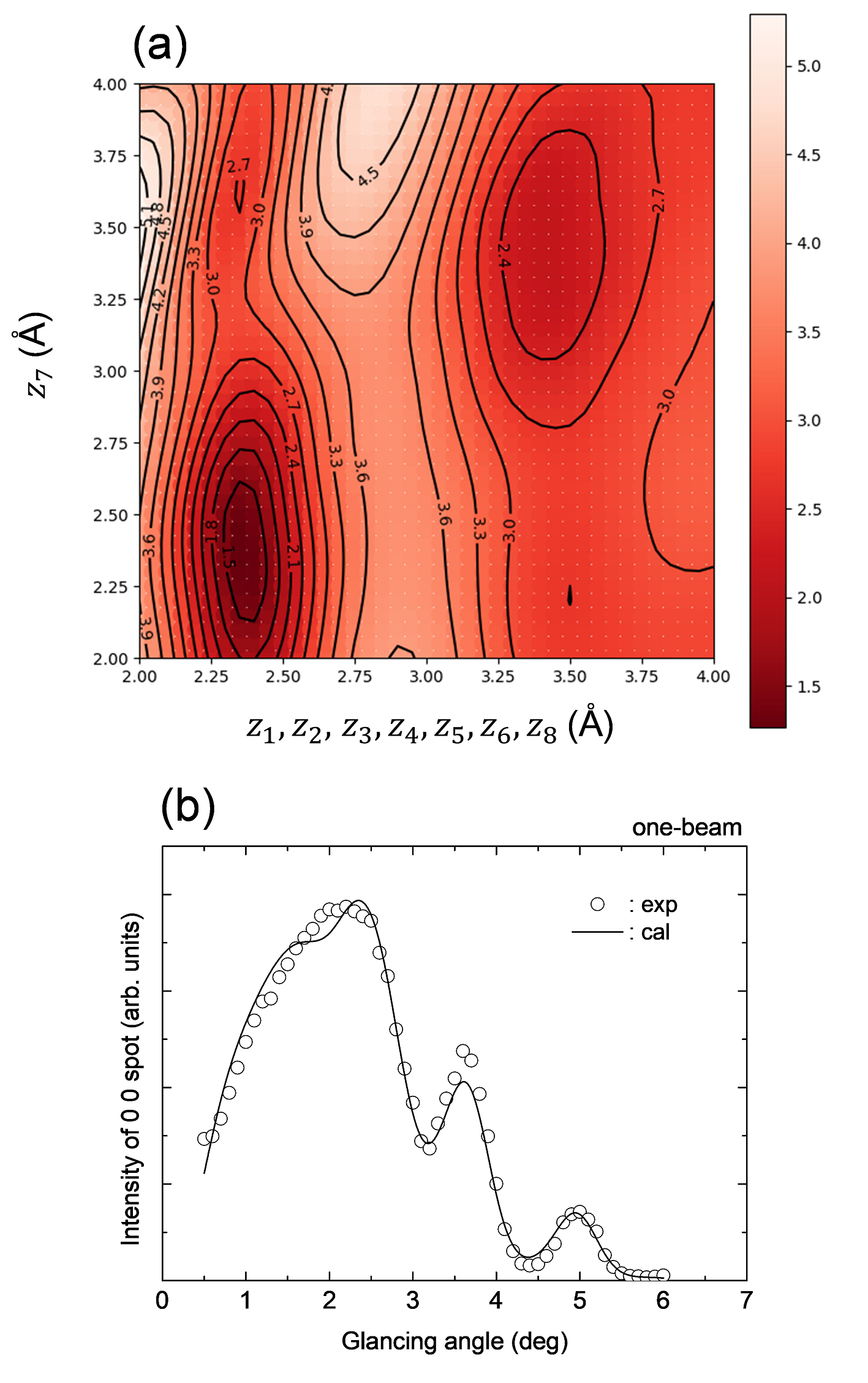}
\caption{(a) A contour plot of the R-factor with respect to z positions of one ($z_7$ in Fig. 2) and the other Si atoms ($z_1$,$z_2$,$z_3$,$z_4$,$z_5$,$z_6$,$z_8$ in Fig. 2) in the Si honeycomb lattice layer on Al(111). (b) Total-reflection high-energy positron diffraction rocking curves for the Al(111)3$\times$3-Si surface, acquired at room temperature with a 10 keV positron beam under the one-beam condition.}
\label{f4}
\end{figure}

Figure 4(a) shows the 2D R-factor results for TRHEPD rocking curves under the one-beam condition [Fig. 4(b)] for the silicene model on Al(111), calculated via massively parallel data-analysis using 2DMAT\cite{Motoyama2022}. In the model, there were eight atoms in the 3$\times$3 unit cell and the height was $z_{i}$ for the $i^{th}$ atom ($i \rm = 1 \sim 8$). To examine deformation of the flat structure, the $z_{i}$ parameters were classified into two groups, one atom ($z_7$) and the other containing atoms with the same height ($z_1$=$z_2$=$z_3$=$z_4$=$z_5$=$z_6$=$z_8$). R-factor values were then plotted with respect to these heights [Fig. 4(a)]. In the 2D diagram, the minimum R-factor of 1.26 $\rm \%$ was found at $z_7$ = 2.40 $\rm \AA$ and $z_1$=$z_2$=$z_3$=$z_4$=$z_5$=$z_6$=$z_8$ = 2.35 $\rm \AA$. The heights, $z_{i}$'s, were essentially equal at each Si site. Further searches for optimum values around the minimum of R-factors were performed using Nelder-Mead methods with 2DMAT\cite{Motoyama2022}. The optimum silicene atomic positions had nearly the same heights ($z_7$ = 2.38 $\rm \AA$ and $z_1$=$z_2$=$z_3$=$z_4$=$z_5$=$z_6$=$z_8$ = 2.37 $\rm \AA$). As shown in Fig. 4(b), the TRHEPD rocking curve calculated from the optimum heights was in good agreement with the experimental curve. Consequently, the analysis confirmed that the silicene layer was flat with no buckling structure.

\begin{figure}[htp]
\centering
\includegraphics[width=8 cm]{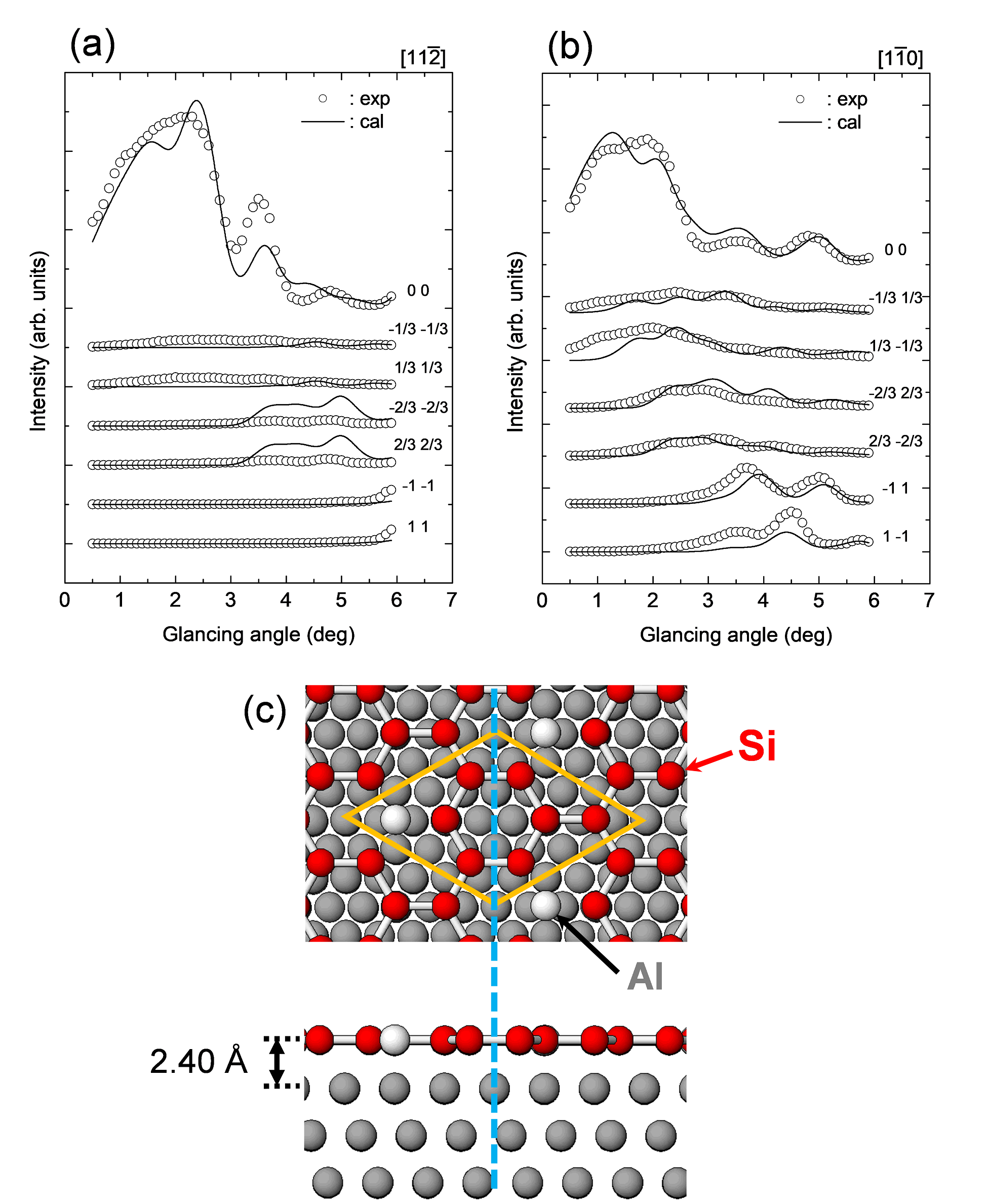}
\caption{Positron diffraction analysis with the Al-embedded silicene model. (a,b) Total-reflection high-energy positron diffraction rocking curves of the Al(111)3$\times$3-Si surface, acquired at room temperature with an 10 keV positron beam incident along the (a) [$11\bar{2}$] and (b) [$1\bar{1}$0] directions. (c) Schematic of the model structure. A 3$\times$3 unit cell is depicted.}
\label{f5}
\end{figure}

Figure 5(a,b) is a set of TRHEPD rocking curves of the Al(111)3$\times$3-Si surface, with comparisons to simulation curves ($z_{1-8}$ = 2.40 $\rm \AA$) of the flat Al-embedded silicene model that replaces one of the Si atoms with an Al atom  [Fig. 5 (c)]. The results show good agreements that confirm the Al-Si mixed model. A R-factor value of the Al-embedded model was R = 2.12$\rm \%$, which was slightly smaller than R = 2.38$\rm \%$ for the flat silicene model (Fig. 2). The R-factor values derived from Figs.4 and 5 were different because the former TRHEPD results were obtained with only a specular beam, while the latter were obtained with many diffracted beams. Previous reports proposed a "kagome-like silicene" model as the surface structure\cite{Sassa2020}. However, the TRHEPD analysis here was in disagreement, as shown in Fig. 12 in the Appendix.

The first-principles calculation with OpenMX has also resulted in the flat structure for the Al-embedded silicene layer on Al(111) after the optimization. A standard derivation of the height (SDH) of the Al-embedded silicene layer on the Al(111) was only 0.06 $\rm \AA$. The result is consistent to the flat surface structure, as determined by the TRHEPD analyses.  


\subsection{Core-level spectra}
Positron diffraction experiments revealed that the Al(111)3$\times$3-Si surface structure could be described by a flat honeycomb lattice model. However, substitution of Al atoms in the Si overlayer was uncertain because of indistinguishable positron scattering. The Si atoms could be bonded with neighboring Si and/or Al atoms. This creates different chemical environments for these Si sites that should result in corresponding chemical shifts in Si core-level photoemission spectra\cite{MatsudaTextbookp1}. 

Figure 6(a) shows Si 2$p$ core-level photoemission spectra of the Al(111)3$\times$3-Si surface. There were at least five peaks, as reported previously\cite{Sassa2020}. Because a single chemical site for Si generates a doublet state, Si 2$p_{3/2}$ and 2$p_{1/2}$, from spin-orbit splitting in the binding energy range, the spectrum indicated the existence of multiple chemical environments for Si atoms at the surface. To reveal the components, spectral curve-fittings were conducted using Doniach-Sunjic functions for various emission angles, $\rm \theta$'s. The spin-orbit splitting and the branching ratio were fixed at 0.4 eV and 0.5, respectively. The Doniach-Sunjic width was fixed at 0.0425 eV to be consistent to the 0.085 eV reference Lorentzian width reported previously\cite{Karlsson1994}. The curve-fit spectra are shown in Fig. 6(b-d), with parameters given in Table I. The results indicated four Si components, or four chemical sites, on the Al(111)3$\times$3-Si surface.

\begin{figure}[htp]
\centering
\includegraphics[width=8 cm]{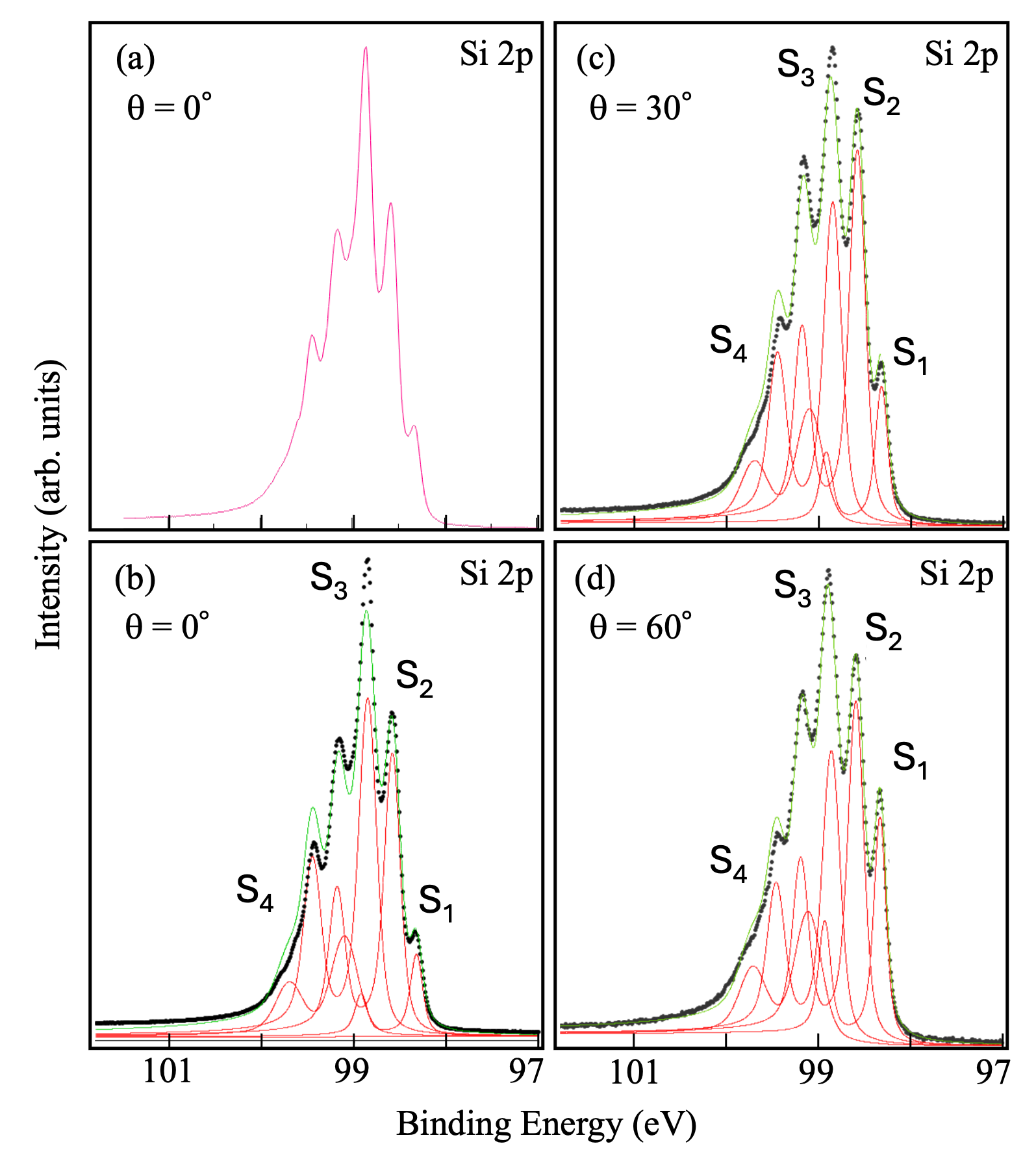}
\caption{Si 2$p$ core-level photoemission spectra of the Al(111)3$\times$3-Si surface, taken at emission angles of (a,b) $\rm \theta = 0^\circ$, (c)$\rm \theta = 30^\circ$, and (d)$\rm \theta = 60^\circ$. The spectra were acquired at room temperature with 150 eV photons. In (b-d), experimental spectra are dotted lines, while curve-fit spectra of the Si components are given by green and red curves, respectively. The curve-fit parameters are summarized in Table I.}
\label{f6}
\end{figure}

\begin{table*}[htp]
  \caption{Fitting parameters of the Si 2$p$ core-level spectra for the Al(111)3$\times$3-Si surface, shown in Fig. 6. The spin-orbit splitting, the Doniach-Sunjic width, and the branching ratio were set at 0.6 eV, 0.0425 eV, and 0.5, respectively.}
  \label{table1}
  \centering
  \begin{tabular}{|c|c|c|c|c|}
    \hline
Components  & $\rm S_1$  & $\rm S_2$ & $\rm S_3$ & $\rm S_4$\\
    \hline 
Binding energy (Si 2$p_{3/2}$) & 98.32 eV & 98.58 eV & 98.85 eV & 99.10 eV\\
Gaussian width & 0.103 eV & 0.151 eV & 0.174 eV & 0.288 eV \\ 
    \hline
  \end{tabular}
\end{table*}

\begin{figure}[htp]
\centering
\includegraphics[width=8 cm]{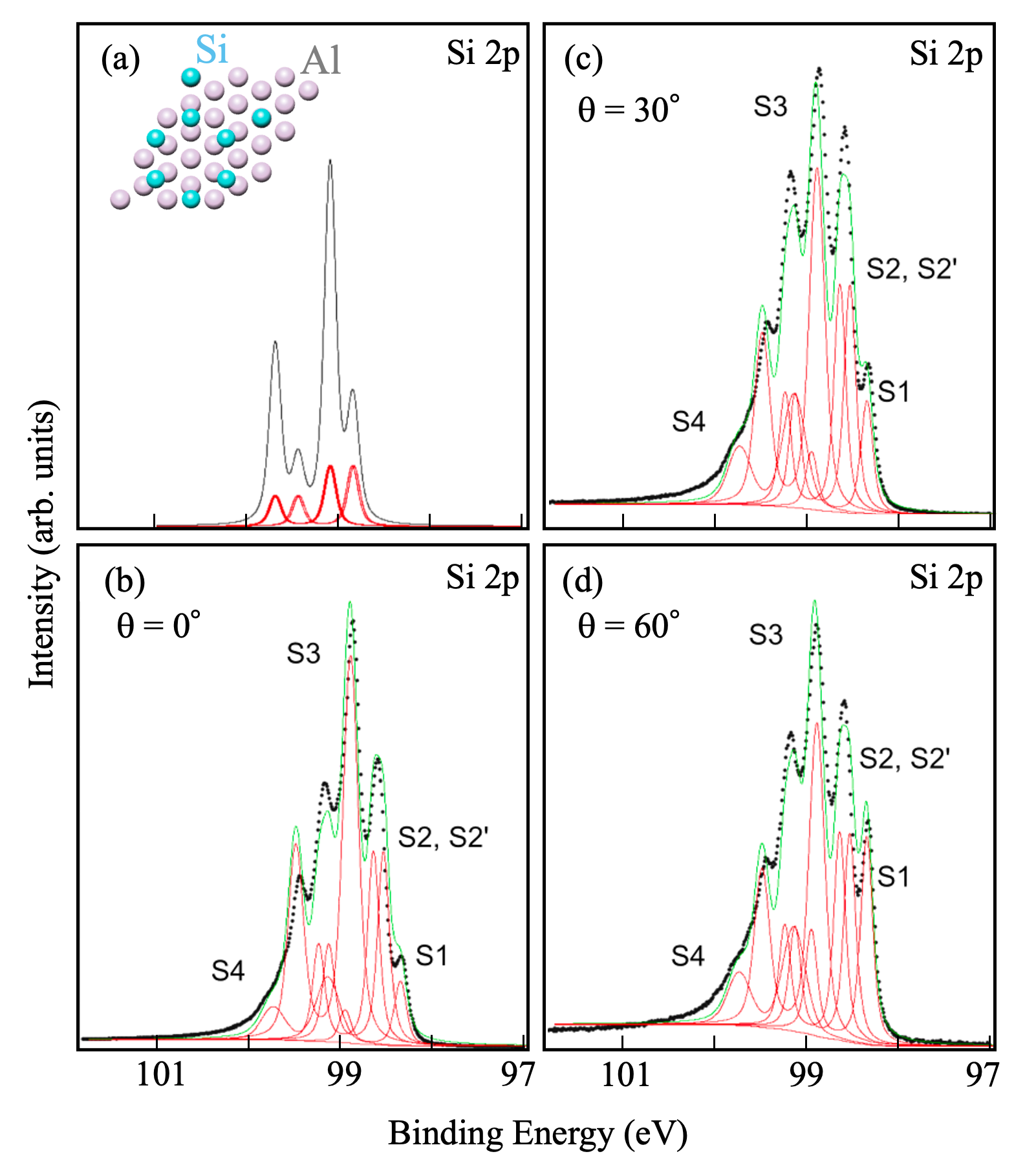}
\caption{Simulations of Si 2$p$ core-level photoemission spectra of surface layers on Al(111). (a) The flat silicene model, and the Al-embedded silicene model for (b) $\rm \theta = 0^\circ$, (c) $\rm \theta = 30^\circ$, and (d) $\rm \theta = 60^\circ$. In (a), the simulated component spectrum is depicted in black (red). The model structure is depicted in the inset. In (b-d), experimental spectra are shown by dotted lines, while simulated spectra and the components are given by green and red curves, respectively. The simulation parameters are summarized in Table III.
}
\label{f7}
\end{figure}

\begin{table*}[htp]
  \caption{Simulated Si 2$p_{3/2}$ binding energies of the Al-embedded silicene models in Fig. 8. Results of the silicene model are shown in Fig. 7(a).}
  \label{table2}
  \centering
  \begin{tabular}{|c|c|c|c|c|}
    \hline
Components & Si(-Si) & Si(-Al) & Si(-Al) & Si(-Al) \\
    \hline 
A number of the NN Al atoms  & 0 & 1 & 2 & 3 \\
    \hline 
Model Fig.8(a) & - & 98.72 eV & - & - \\
Model Fig.8(b) & 98.79 eV & - & - & - \\
Model Fig.8(c) & - & 98.70 eV & - & - \\
Model Fig.8(d) & - & - & 98.57 eV & - \\
Model Fig.8(e) & 98.74 eV & - & - & - \\
Model Fig.8(f) & - & 98.69 eV & - & - \\
Model Fig.8(g) & - & - & 98.60 eV & - \\
Model Fig.8(h) & - & - & - & 98.41 eV \\
Silicene (atop site)  & 98.83 eV & - & - & - \\
Silicene (bridge site)  & 99.08 eV & - & - & - \\
    \hline
  \end{tabular}
\end{table*}

\begin{table*}[htp]
  \caption{Simulation parameters for the Si 2$p$ core-level spectra of the Al(111)3$\times$3-Si surface, shown in Fig. 7. The spectral components were given by Voigt functions with 0.085 eV Lorentzian widths and Gaussian widths given in the table. Spin-orbit splitting and a branching ratio were set at 0.6 eV and 0.5, respectively. The simulation was performed using OpenMX\cite{Ozaki2017,Ozaki2003}.}
  \label{table3}
  \centering
  \begin{tabular}{|c|c|c|c|c|c|}
    \hline
Components  & S1  & S2 & S2' & S3 & S4\\
    \hline 
Binding energy (Si 2$p_{3/2}$) & 98.41 eV & 98.59 eV & 98.70 eV & 98.95 eV & 99.2 eV\\
Gaussian width & 0.10 eV & 0.10 eV & 0.10 eV & 0.15 eV & 0.15 eV\\
Spectral weight ($\rm \theta = 0^\circ$) & 5.9 $\%$ & 17.6 $\%$ & 17.6 $\%$ & 47.1 $\%$ & 11.8 $\%$\\
Spectral weight ($\rm \theta = 30^\circ$) & 9.1 $\%$ & 18.2 $\%$ & 18.2 $\%$ & 36.4 $\%$ & 18.2 $\%$\\
Spectral weight ($\rm \theta = 60^\circ$) & 16.7 $\%$ & 16.7 $\%$ & 16.7 $\%$ & 33.3 $\%$ & 16.7 $\%$\\
    \hline
  \end{tabular}
\end{table*}


\begin{figure}[htp]
\centering
\includegraphics[width=8 cm]{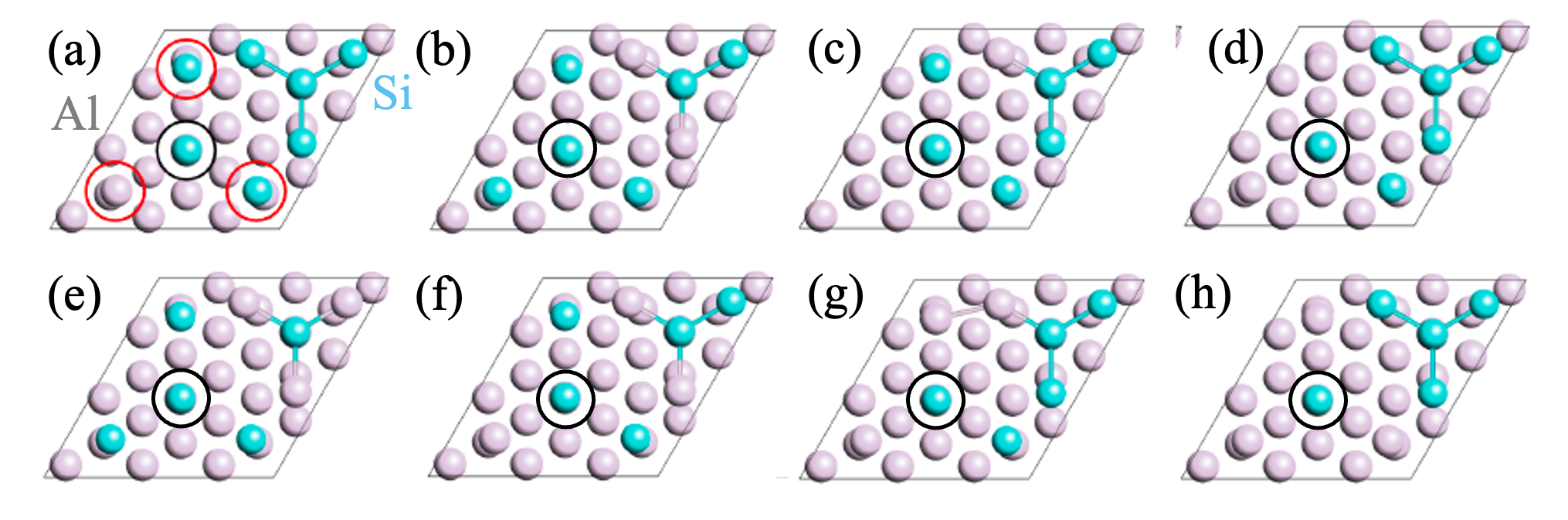}
\caption{Structural models of Al-embedded silicene layers with Si coverages of (a) 7/9 ML, (b-d) 6/9 ML, and (e-h) 5/9 ML. Black circles indicate Si atoms used for calculating the Si 2$p_{3/2}$ binding energies, while the red circles in (a) indicate example positions of nearest-neighbor atoms in the surface layer.
}
\label{f9}
\end{figure}

To analyze chemical environments at the Al(111)3$\times$3-Si surface, the first-principles spectral simulations were calculated using OpenMX\cite{Ozaki2003,Ozaki2004,Ozaki2005,Lejaeghere2016,Ozaki2017,OPENMX}. Figure 7(a) shows four prominent peaks in a simulated spectrum for the silicene layer on Al(111). Because of spin-orbit splitting, each Si 2$p$ core-level appeared as a doublet state, and the simulations indicated two Si components for the model structure. As depicted in the inset, one was attributed to Si atoms at atop Al sites and the other to bridge Al sites. While the calculations successfully linked the surface structure and the core-level spectra, the simulated results for the silicene layer were completely different from the experimental data (Fig. 6). Thus, it was necessity to consider a surface structure beyond a simple silicene layer for the Al(111)3$\times$3-Si surface.

To reproduce the experimental results, the spectral simulations were performed for various Si sites in the Al-embedded silicene layer model (Fig. 8). As noted earlier, Si atoms in the surface layer can bond with different numbers of intralayer Si and/or Al atoms. For example, in Fig. 8(a), one Si atom used for the calculations was bonded with two Si atoms and one Al nearest neighbor (NN). Table II lists binding energies of the Si 2$p_{3/2}$ level for various structure models of silicene and Al-embedded silicene layers on Al(111). The core-level binding energies varied with respect to the number of NN Al atoms in the Si-Al bonding. Smaller NN values had higher binding energies, which was consistent with the silicene model results given in Table II. Moreover, there was a quantitative relationship between the binding energy and the NN value: $\rm 98.4 \pm 0.1 eV$ (NN=3), $\rm 98.6 \pm 0.1 eV$ (NN=2), $\rm 98.7 \pm 0.1 eV$ (NN=1), and $\rm 98.8 \pm 0.1 eV$ (NN=0). Because the model calculations in Fig. 8 considered only a single (atop) site on the Al substrate, differences in the substitution sites should be considered in an energy range ($\rm \pm 0.1 eV$), as in the calculation results of the silicene model (Table II). 

On the basis of the calculated binding energies, the experimental spectra of the Al(111)3$\times$3-Si surface were fairly well reproduced by five components, S1, S2, S2', S3 and S4, with the parameters in Table III. From the binding energies, S3 and S4 could be assigned to Si atoms that bonded only with NN Si atoms, while S1, S2, and S2' were also bonded to NN Al atoms. A ratio of these two types of components was used to evaluate the extent of Al substitutions in the Al-embedded silicene layer. Those ratios, $R_{Al/Si}$'s, reflecting numbers of Si atoms bonding with Al atoms (S1, S2, S2' components) and those without Al bonding (S3, S4 components), were evaluated from Table III. Specifically, $R_{Al/Si}$ = 0.7, 0.83, and 1 for spectra acquired at $\rm \theta = 0^\circ$, $\rm 30^\circ$, and $\rm 60^\circ$, respectively, in Figs. 6 and 7. According to previous reports on Al(111)3$\times$3-Si, the 2D phase was found at Si coverages from 4/9 ML to 7/9 ML\cite{Munoz1,Munoz2,Jona,Sassa2020}. $R_{Al/Si}$ $\rm \leq$ 1 could only be satisfied at 7/9 ML coverage with $R_{Al/Si}$ = 0.75. For the eight sites in the honeycomb lattice in a 3$\times$3 unit cell, this result corresponded to one Al-substitution. The Al atom bonded with three Si atoms at NN sites (S1, S2, S2' components) and the remaining four Si atoms bonded only with NN Si atoms (S3, S4 components). The Al-embedded silicene model corresponds to that in Fig. 5(c), and it reproduced the TRHEPD curves, as shown in Fig. 5(a,b). There were slight deviations between experiments and simulations for photoemission binding energies and for intensities in positron diffraction rocking curves. This likely indicated that embedded Al atom positions were not specific but rather random in the eight sites in the 3$\times$3 unit cell. Thus, the structure of the Al(111)3$\times$3-Si surface could be described as a layer of Al-embedded silicene with a Si coverage of 7/9 ML.


\begin{figure}[htp]
\centering
\includegraphics[width=7 cm]{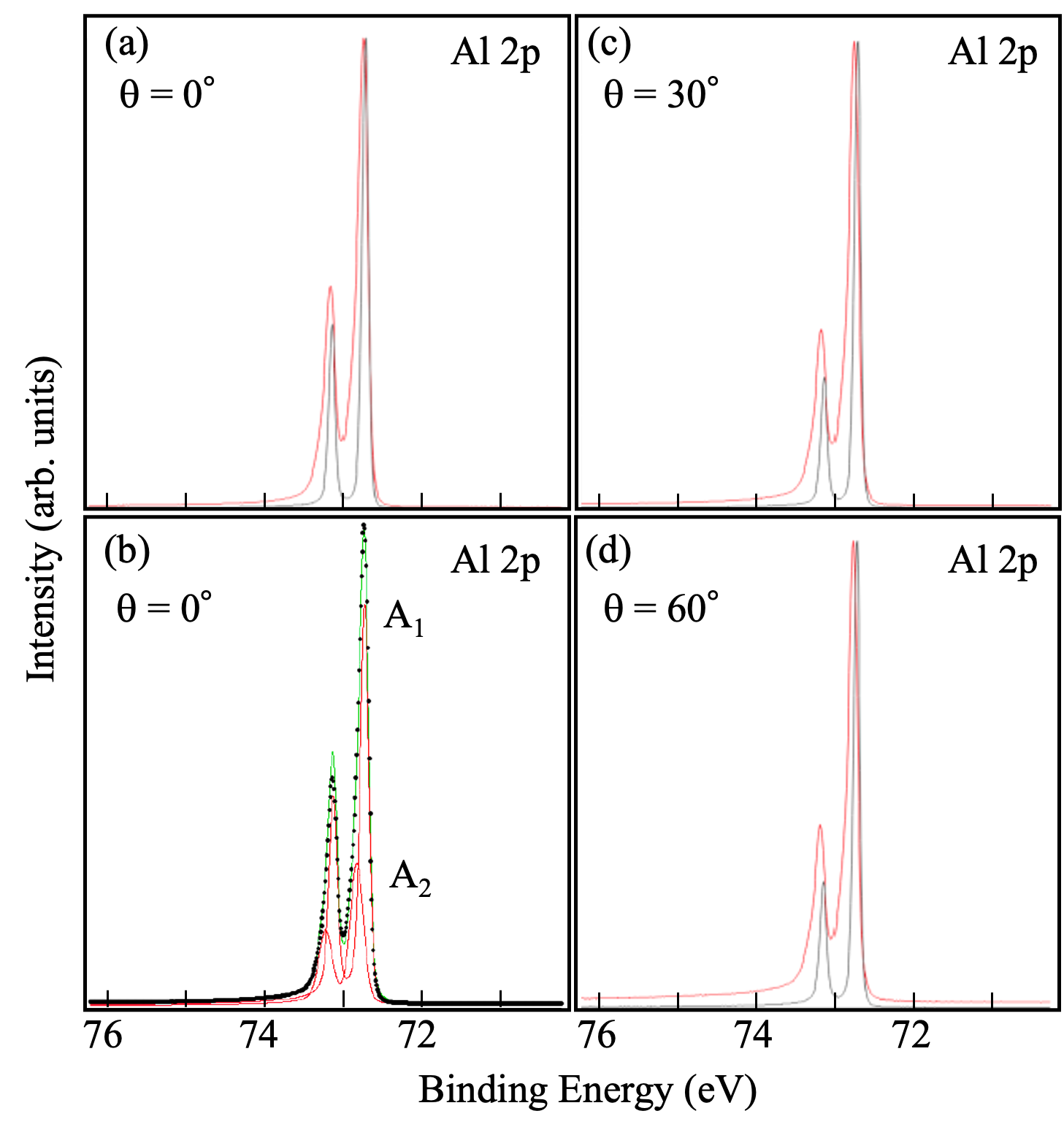}
\caption{Al 2$p$ core-level photoemission spectra of the Al(111)3$\times$3-Si surface, acquired at emission angles (a,b) $\rm \theta = 0^\circ$, (c)$\rm \theta = 30^\circ$, and (d)$\rm \theta = 60^\circ$. The spectra were acquired at room temperature with an incident photon energy of h$\rm\nu$=123 eV. In (a,c,d), red and black solid curves correspond to the Al(111)3$\times$3-Si and the Al(111) surfaces, respectively. In (b), an experimental spectrum of (a) is shown by a dotted line, while the curve-fit spectrum and its components are given by green and red curves, respectively. The fitting parameters are summarized in Table IV.}
\label{f2}
\end{figure}
 
\begin{table}[htp]
  \caption{Fitting parameters for the Al 2$p$ core-level spectra of the Al(111)3$\times$3-Si surface, shown in Fig. 2(b). Spin-orbit splitting, the Doniach-Sunjic width, and the branching ratio were set at 0.4 eV, 0.015 eV, and 0.5, respectively.}
  \label{table1}
  \centering
  \begin{tabular}{|c|c|c|}
    \hline
Components  & $A_{1}$  & $A_{2}$ \\
    \hline 
Binding energy (Al 2$p_{3/2}$) & 72.73 eV & 72.83 eV\\
Gaussian width & 0.107 eV & 0.176 eV \\ 
    \hline
  \end{tabular}
\end{table}

To examine the chemical environment at the Al site, a set of Al 2$p$ core-level spectra for the pristine Al(111) and the Al(111)3$\times$3-Si surfaces were acquired, as shown in Fig. 9(a). The Al core-level states generated a doublet structure via spin-orbit splitting, and a pair of the 2$p_{3/2}$ and 2$p_{1/2}$ peaks shows broadening at higher binding energies upon formation of the 3$\times$3 surface superstructure. The spectral broadening was confirmed in the data taken at different emission angles, as shown in Fig. 9(c,d). To examine the spectral features in detail, curve-fitting was performed for the Al(111)3$\times$3-Si spectrum, which revealed two components at binding energies of 72.73 eV ($A_1$) and 72.83 eV ($A_2$), as shown in Fig. 9(b). The fit used parameters that are summarized in Table IV. Because the $A_1$ component was also observed on the pristine Al(111) surface, it could be assigned to bulk Al. However, the $A_2$ component was assigned to surface Al atoms, and likely confirmed the presence of an Al atom in the Al-embedded silicene layer. Similar spectral broadening was reported for Ag 3$d$ core-level spectra of the Ag(111)4$\times$4-Si surface that had a silicene overlayer\cite{Kuuchle2022}. The $A_2$ peak may contain contributions from Al atoms in the Al-embedded silicene layer and those in the subsurface.

\begin{figure}[htp]
\centering
\includegraphics[width= 6 cm]{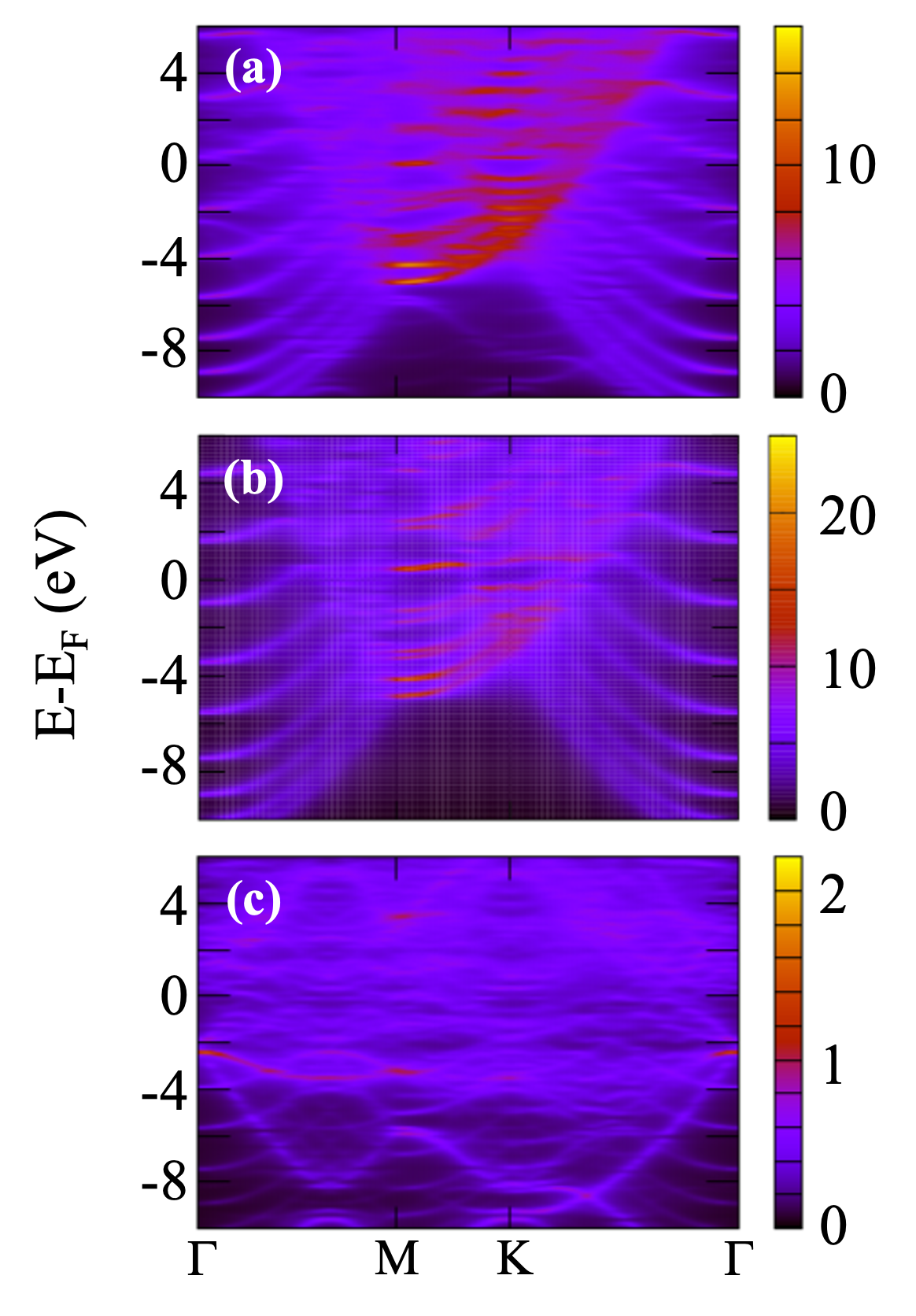}
\caption{Calculated band structure of the Al-embedded silicene model on Al(111). (a) Total band diagrams, including both surface and substrate contributions. (b) Band diagram of the Al(111) substrate. (c) Band diagram of a Al-embedded silicene surface layer at a Si coverage of 7/9 ML. Both band structures were unfolded into the Al(111)1$\times$1 surface Brillouin zone with labeled high symmetry points, $\rm \Gamma$, M, and K. The color scale corresponds to the spectral weights of the electronic states.
}
\label{f10}
\end{figure}

\begin{figure}[htp]
\centering
\includegraphics[width=8 cm]{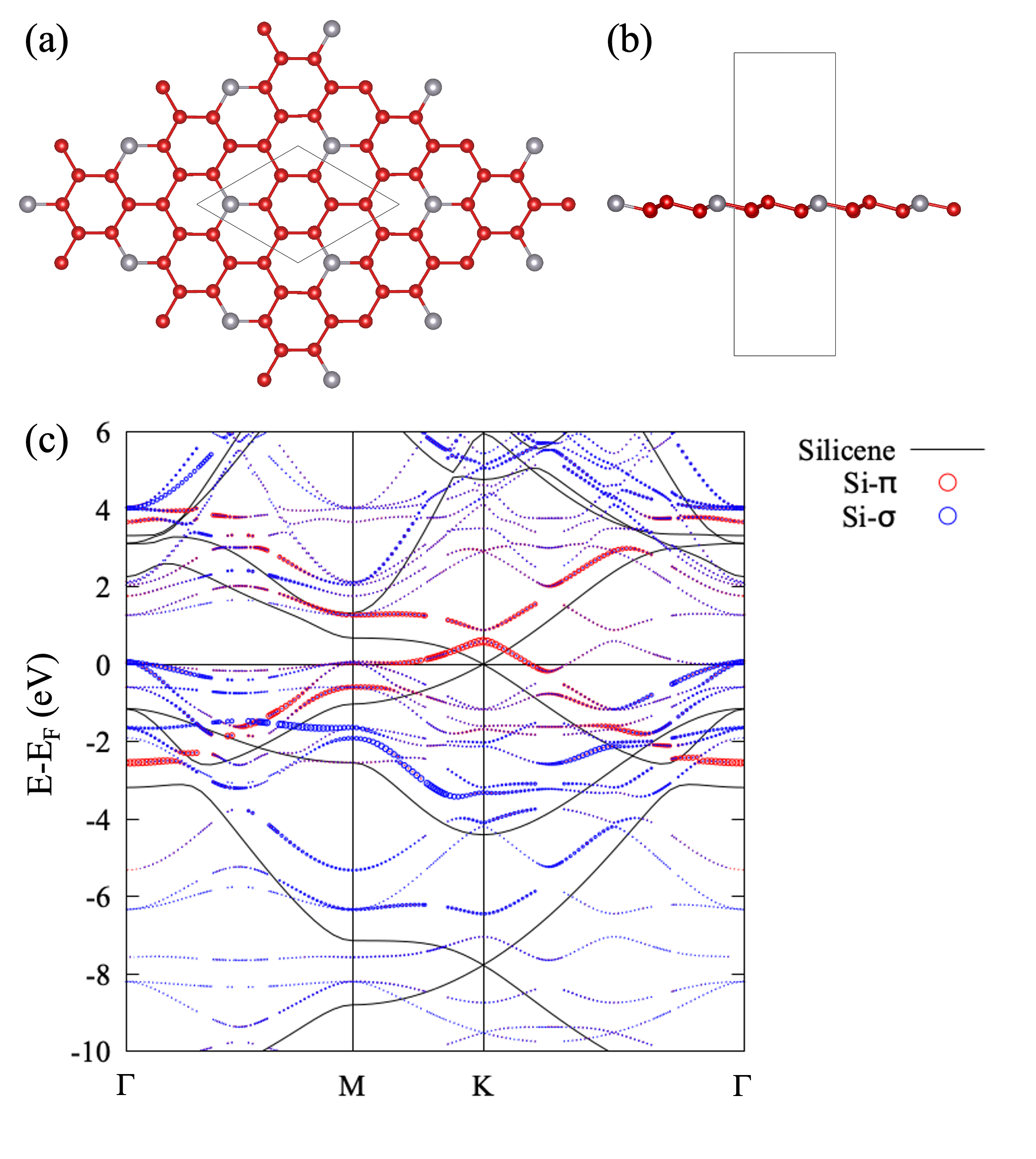}
\caption{Electronic structure of a free-standing layer of Al-embedded silicene. Optimized structure model: (a) top and (b) side views. Silicon and aluminum atoms are shown as red and gray circles, respectively. A unit cell for the calculations is depicted in the figure. (c) Calculated band structure of the free-standing atomic layers: (blue, red) Al-embedded silicene and (black) silicene. The red and blue color scale corresponds to the spectral weights of electronic states in the Si-$\pi$ and Si-$\sigma$ bands, respectively. Band-unfolding was applied to compare band diagrams of the layers in the same Brillouin zone\cite{Lee2013}.
}
\label{f11}
\end{figure}
  
\subsection{Surface band structure}
The structure model was determined to be the Al-embedded silicene layer on Al(111), as shown in Fig. 5(c), and therefore electronic band structures were calculated for the Al(111)3$\times$3-Si surface. Figure 10 shows the calculated band diagrams with different weights for the surface system: (a) the total system, (b) the Al(111) substrate, and (c) the Al-embedded silicene surface layer. The band structure appeared complex at the Fermi level ($\rm E_F$) because of hybridizations between the substrate and surface electronic states. At the $\rm \Gamma$ point, there are electronic states at $\rm E - E_{F} \sim$ - 2 eV in Fig. 10 that show dispersion curves at higher binding energies along the $\rm \Gamma$-M and $\rm \Gamma$-K directions. The electronic features were found in the experimental band diagram reported previously via angle-resolved photoemission spectroscopy\cite{Sato2020}. These agreements confirmed the present structure model. The calculated band diagram was similar to that reported for a pristine silicene layer on Al(111)\cite{Sato2020}. Information from the band structure was not sufficient to distinguish structure models of silicene and Al-embedded silicene overlayers.

The electronic structure of the Al-embedded silicene layer on Al(111) appeared complex because of interactions between the overlayer and the substrate, as shown in Fig. 10. Thus, it was not suitable to use the system to understand electronic variations of a silicene layer attributed to Al substitution. Instead, we conducted band calculations of a free-standing Al-embedded silicene layer in vacuum, as shown in Fig. 11. After structural optimization, the free-standing layer exhibited a buckled structure with a 0.29 $\rm \AA$ in SDH. In contrast, the OpenMX first-principles calculation indicated that the layer was flat (0.06 $\rm \AA$ in SDH) on the Al(111) substrate. For reference, the SDH of the pure silicene on the Al(111) was 0.02 $\rm \AA$, while the SDH of a pure silicene layer in vacuum was 0.36 $\rm \AA$.

Concerning the band structure, the Si-$\pi$ band formed a gap at the K point for the surface, as opposed to the Dirac cones in the silicene layer shown for comparison. At $E_F$, the Si-$\sigma$ bands dispersed at higher energies and similar curves were found in the surface system [Fig. 10(c)]. The conservation was sharply in contrast to the Si-$\pi$ band at the surface. The difference was likely attributed to layer-substrate interactions along the surface normal, which was sensitive (insensitive) to the Si-$\pi$ (Si-$\sigma$) bands. Such electronic properties were also reported for various surface systems\cite{Mathis}. The Al atom has been reported to be an acceptor-type donor in bulk (diamond) silicon. In the Al-embedded silicene layer, the Mulliken charge of the Si atoms was 3.89 $\sim$ 3.94 $e^{-}$, while that of the Al atoms was 3.01 $\sim$ 3.06 $e^{-}$, where neutral Si and Al atoms in the calculations have electrons of 4 and 3 $e^{-}$, respectively. A substituted Al atom in the silicene layer was essentially neutral and was not suitable as an acceptor-type donor.

\section{Conclusion}
The surface structure of a long-range ordered 3$\times$3 phase, prepared by Si deposition on Al(111), was subjected to a multi-beam investigation using positron diffraction and X-ray photoemission spectroscopy. Total reflection high-energy positron diffraction analyses revealed that the surface layer had a flat honeycomb lattice structure. Core-level photoemission spectroscopy indicated that eight sites in the 3$\times$3 unit cell were occupied by one Al atom and seven Si atoms. Hence, the superstructure could be described as a Al-embedded silicene layer on the Al(111) surface. This work thus revealed a silicene layer that contained a substrate atom. This could provide a future approach to dope external atoms in functional Xene surface layers.

\begin{acknowledgments}
This work was supported by JSPS KAKENHI grants (Grants No. JP21H05012, No. JP19H04398, and No. JP18H03874), by a Grant-in-Aid for JSPS Fellows (Grant No. 21J21993), and by JST, CREST Grant No. JPMJCR21O4, Japan. The preliminary experiment was performed at facilities of the Synchrotron Radiation Research Organization, the University of Tokyo. The parts of this work was performed under the approval of the Photon Factory Program Advisory Committee (Proposal No. 2014S2-004). The numerical computation was carried out partially on the supercomputers of the Supercomputer Center, Institute for Solid State Physics, University of Tokyo. The numerical computation was carried out partially, also, on the Fugaku supercomputer through the HPCI projects (hp210228, hp210267, hp220248, hp230304) and the Wisteria-Odyssey supercomputer as part of the Interdisciplinary Computational Science Program in the Center for Computational Sciences, University of Tsukuba. We acknowledge Elettra Sincrotrone Trieste for providing access to its synchrotron radiation facilities and we thank Dr. Paolo Moras, Dr. Polina M. Sheverdyaeva, and Dr. Asish Kumar Kundu for assistance in using the VUV-Photoemission beamline (beamtime nr. 20185157). A. K. K. acknowledges the project EUROFEL-ROADMAP ESFRI of the Italian Ministry of University and Research.
\end{acknowledgments}

\appendix*
\section{TRHEPD analysis of the kagome-like silicene model}
Figure 12(a,b) is a set of TRHEPD rocking curves that are compared with a simulation of the kagome-like silicene model\cite{Sassa2020} depicted in Fig. 12(c). The structure contained various types of Si atoms that generate the kagome network on the Al(111) substrate and behaved as adatoms. The model was proposed to qualitatively explain the complex Si 2$p$ core-level spectra of the surface that indicated existence of various Si sites\cite{Sassa2020}. As shown in Fig. 12(a,b), the correlation between experimental and simulated curves was poor (R = 3.65$\rm \%$) relative to that for the Al-embedded silicene model (R = 2.12$\rm \%$) in Fig. 5. Thus, the Al-embedded silicene model was more likely than the kagome-like silicene model for the structure of the Al(111)3$\times$3-Si surface.

\begin{figure}[htp]
\centering
\includegraphics[width=8 cm]{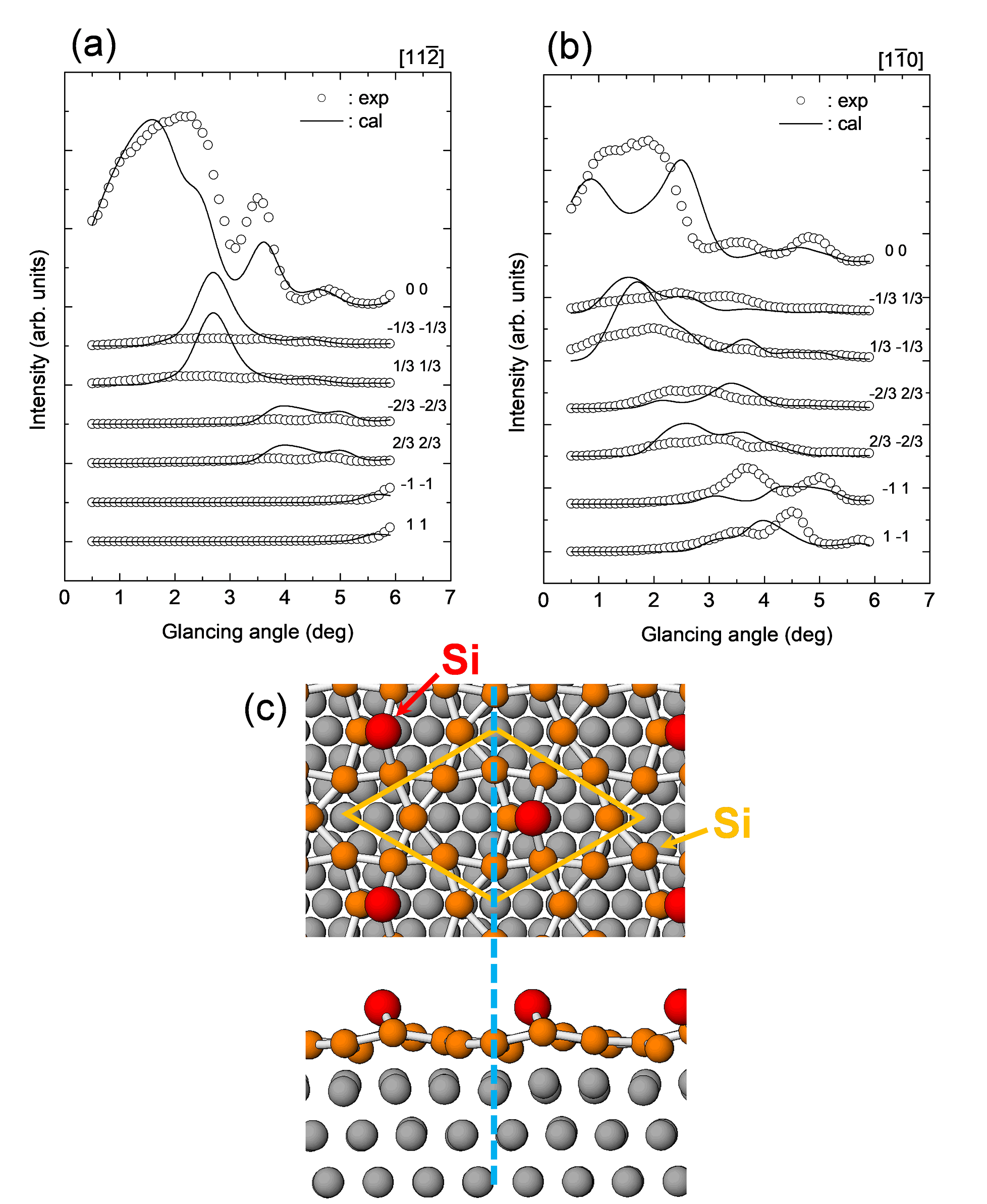}
\caption{Positron diffraction analysis of the kagome silicene model. (a,b) Total-reflection high-energy positron diffraction rocking curves of the Al(111)3$\times$3-Si surface, acquired at room temperature with a 10 keV positron beam incident along the (a) [$11\bar{2}$] and (b) [$1\bar{1}$0] directions.(c) Schematic of the structure model.}
\label{f12}
\end{figure}

\end{document}